 \documentclass[prb,aps,twocolumn]{revtex4} 
\usepackage[dvipdfmx]{graphicx} 
\usepackage{tabularx}
\usepackage{dcolumn} 
\usepackage{color}
\usepackage{bm}
\usepackage{amsmath}
\usepackage{amssymb}
\usepackage{here}
\usepackage{ulem}

\begin{document} 


\title{Anisotropic field response of specific heat for a ferromagnetic superconductor UCoGe in magnetic fields}

\author{Shota Nakamura$^{1,2}$, Shunichiro Kittaka$^{2,3}$, Kazushige Machida$^4$, Yusei Shimizu$^{2,5}$, Ai Nakamura$^5$,
 Dai Aoki$^5$,  Toshiro Sakakibara$^2$}
\affiliation{$^1$Nagoya Institute of Technology, Aichi, Nagoya 466-8555, Japan}
\affiliation{$^2$Institute for Solid State Physics, The University of Tokyo,  Kashiwa 277-8581, Japan}  
\affiliation{$^3$Department of Physics, Faculty of Science and Engineering, Chuo University, Kasuga, Bunkyo-ku, Tokyo 112-8551, Japan} 
\affiliation{$^4$Department of Physics, Ritsumeikan University, Kusatsu 525-8577, Japan}
\affiliation{$^5$Institute for Materials Research, Tohoku University, Oarai 311-1313, Japan; Japan Atomic Energy Agency, Tokai 319-1106, Japan} 

\date{\today}

\begin{abstract}
Magnetic-field-angle-resolved specific heat and magnetization measurements were conducted on a ferromagnetic superconductor UCoGe with remarkable anisotropic upper critical field $H_{\rm c2}$. Although $H_{\rm c2}$ reaches a high magnetic field ($\sim 20$~T) along the $b$ axis, it is small ($\sim~0.6$~T) when a magnetic field is applied along the magnetic easy $c$-axis. This study indicates that the specific heat is abruptly suppressed  when the magnetic field is applied toward the $c$ axis from the $a$ and $b$ axes in the ferromagnetic state.
The field response of density of states (DOS) is anisotropic, relative to the $c$ axis, and its angle dependence is slightly singular.
The Ising-type magnetic anisotropy of the ferromagnetic state is dominant even in the anisotropic reinforced superconducting state. These facts indicate that the suppression of DOS may closely relate to the superconducting state. We theoretically analyze these findings together with
URhGe and UTe$_2$ by highlighting the common and distinctive features among three compounds. 
\end{abstract}

\maketitle 

\maketitle

\section{Introduction}
The coexistence of ferromagnetism and superconductivity has been theoretically and experimentally studied since the 1950s.
For a time ferromagnetism was believed to compete with superconductivity, as the large internal magnetic field due to the ferromagnetic order destroys the Cooper pairs with the spin-singlet state.
In fact, electrons that trigger superconductivity differ in orbital
characteristics and constituent atomic species from those that play ferromagnetic (FM) roles in (Ce$_{\rm 1-x}$Gd$_{\rm x}$)Ru$_2$, RuSr$_2$GdCu$_2$O$_8$, ErRh$_4$B$_4$, and HoMo$_6$S$_8$, which are known as FM superconductors~\cite{kitaoka,bernhard,fertig,lynn,review}.

However, a result in contrast to the conventional was reported by Saxena et al.~\cite{saxena} in 2000.
They found that ferromagnet UGe$_2$ under Curie temperature  $T_C$~=~52~K shows superconductivity at $T_{\rm SC}$~=~0.8 K under pressure in the ferromagnetic state~\cite{saxena}.
The discovery of the uranium-based FM superconductor UGe$_2$ is significant, because ferromagnetism and superconductivity are thought to originate from the same $5f$ electrons of uranium in this material.
Subsequently discovered uranium-based FM superconductors URhGe~\cite{aoki} and UCoGe~\cite{huy} exhibit superconductivity in the FM state at ambient pressure.
In a recently discovered heavy fermion superconductor UTe$_2$, superconducting transition occurs from the paramagnetic state, which was thought to be at the proximity of the ferromagnetic order~\cite{ran,UTe2}. This stimulated renewed interest in these FM superconductors. However, the situation is more complicated in UTe$_2$.

These uranium-based superconductors demonstrate exceptional superconducting (SC) properties, which include the microscopic coexistence of superconductivity and ferromagnetism~\cite{hattori,ohta,visser,kotegawa}, 
possible occurrence of an odd-parity pairing \cite{saxena,aoki,huy}, 
and the significant enhancement of $H_{\rm c2}$ exceeding the Pauli-limiting field  $H_{\rm P}$ \cite{sheikin,huy,aoki2,levy}.

The value of a superconducting upper critical magnetic field $H_{\rm c2}$ at 0 K highly exceeds  that of $H_{\rm P}$ expected by a SC order temperature $T_{\rm SC}$ at zero field, $H_{\rm P}$ = 1.86$T_{\rm SC}$ on the basis of the weak coupling BCS model, thus suggesting the spin triplet state of UCoGe.
These anomalous behaviors appear around the FM quantum phase transition; hence, magnetic quantum fluctuations are considered to be responsible for the emergence of  such unusual SC states \cite{tokunaga,hattori,taufour}.

UCoGe exhibits the FM order with the small ordered moment $\mu_{\rm 0} \sim 0.05~\mu_{\rm B}$/U with the Curie temperature $T_{\rm C} \sim 3$~K~\cite{huy}, and the itinerant ferromagnetism exhibits Ising-type magnetic anisotropy.
In addition, its magnetization is oriented in the easy  $c$ axis of the orthorhombic TiNiSi-type crystal structure with the space group, $Pnma$~\cite{canepa}. 
A unique spin-triplet SC state is realized below $T_{\rm SC} \sim 0.6$~K in this material, and an anisotropic $H_{\rm c2}$ is observed.
When a magnetic field is applied in the $c$ axis, superconductivity disappears immediately at $H^c_{\rm c2}\sim$0.6T; however, in the $a$ or $b$ axes, the SC state is maintained, up to a higher magnetic field above 15~T~\cite{aoki2}.
The origin of the mysterious anisotropic feature of $H_{\rm c2}$ can be 
related to the pairing symmetry and mechanism of the superconductivity.
The previous angle-dependent nuclear magnetic resonance (NMR) measurement~\cite{hattori,ishida2021} suggests that the NMR relaxation time, $1/T_1$, is suppressed in a magnetic field along the $c$ axis. They interpret it that the longitudinal FM spin fluctuations tuned by $H\parallel c$ induce the unique superconductivity in UCoGe. 

The objectives of this study are to clarify the origin of the singular field-angle-dependent of
DOS structures, both in the FM and SC states on UCoGe, via field-angle-resolved specific heat measurements backed up by magnetization measurements.  In addition, this study attempts to obtain fundamental thermodynamic information on this material in low-temperature conditions, to facilitate the possible nodal gap structure and underlying pairing mechanism.


\section{Experimental Procedure}

A high-quality single crystalline UCoGe was grown at Institute for Materials Research (IMR), Tohoku Univ. and cut into a rectangular parallelepiped with a weight of 5.7 mg. The resistivity shows the large residual resistivity ratio, $RRR$ = 98.
The magnetic-field-angle-resolved specific heat $C(T, H, \phi, \theta_a, \theta_b)$ was measured by the quasi-adiabatic heat-pulse method using a  $^3$He-$^4$He  dilution refrigerator.
To investigate $C(T, H, \phi, \theta_a, \theta_b)$ in high-magnetic field regions up to 14.5~T near the $b$ axis, in-situ field-angle measurements were also conducted using a home-made two-axis rotation device ~\cite{nakamura} (measurement I).
$C(T, H, \phi, \theta_a, \theta_b)$ was obtained in a wide range of angles $-30^\circ \leq \phi \leq 120^\circ$($\phi$: angle from the $a$ axis toward  the $b$ axis) and $0^\circ \leq \theta_{a, b} \leq 100^\circ$ ($\theta_a, \theta_b$ : angle from the $a$ or $b$ axis toward the $c$ axis, respectively) via the field-orientation control system~\cite{kittaka0} in the magnetic field up to 5~T (measurement II).  To prevent the sample from rotating owing to strong magnetic torque, we used a home-made cell in the present specific heat measurements~\cite{kittaka0}.

We note that the addenda heat capacity is approximatively  12 - 30~$\%$ of the sample heat capacity in the present measurement, and it substantially influences the measured specific heats. Because different calorimeters were adopted in measurements I and II, a significant  subtraction error of $\sim 5$~mJ/K$^2$mol exists in the addenda background.
To review and compare with specific heat data, temperature dependence of the magnetization has been measured by using MPMS. 

\section{Experimental results}

\subsection{$T$-dependent specific heats near the $b$ axis}

\begin{figure}
\includegraphics[width=8cm]{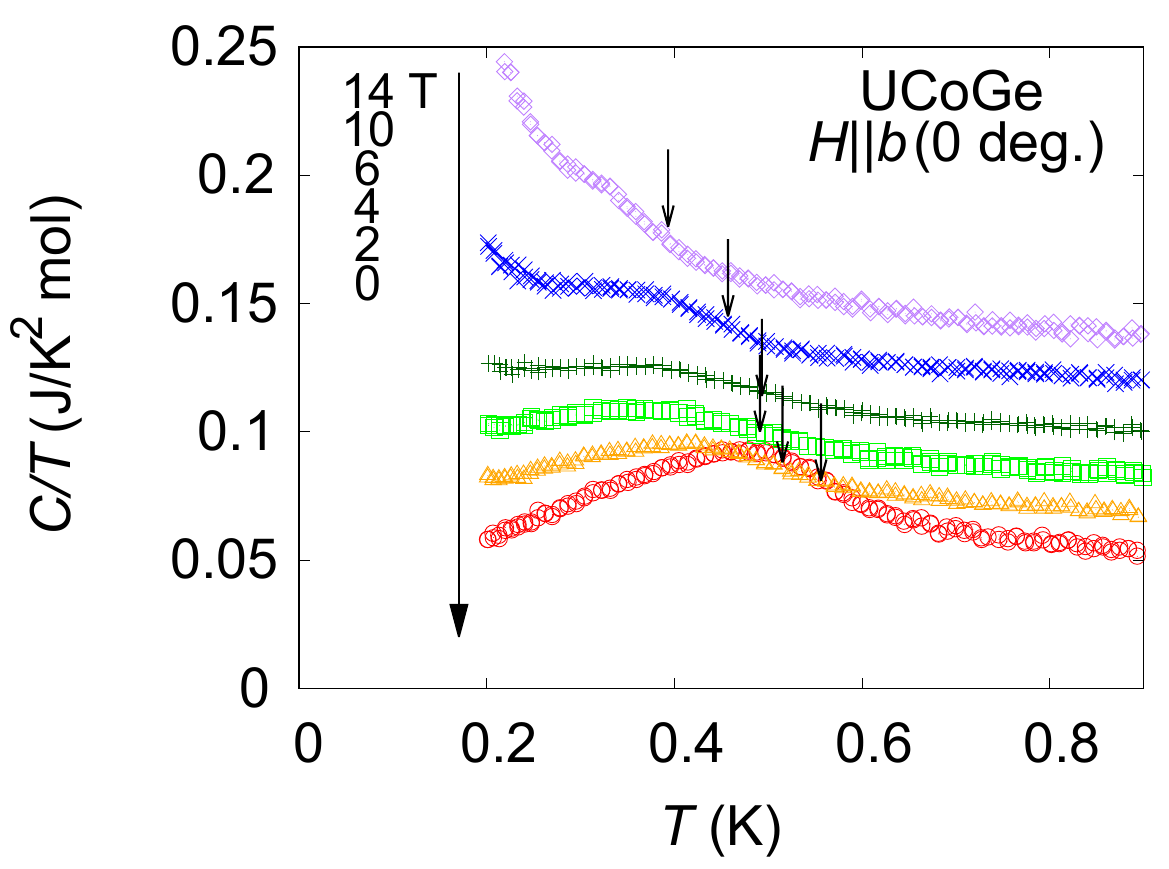}
\caption{(Color Online) Temperature dependence of $C/T$ measured at the several magnetic fields 
along the $b$ axis ($\theta_b = 0$). For clarify, the data are vertically shifted by 0.015. The arrows depict the critical temperature position of the SC state $T_{\rm SC}(H)$. Note the upturns of $C/T$ in $H=10$~T and 14~T at lowest temperatures.}
\label{C-T_b}
\end{figure}

The $H$-$T$ phase diagram of UCoGe near the $b$ axis is obtained through specific heat measurements.
Every $C/T$ value in this study is obtained by subtracting the nuclear term, which is proportional to $T^{-3}$, and we assume that $C/T$ is equal to the electronic specific heat coefficient $\gamma$ in the FM state.
The nuclear term is estimated by using nuclear magnetic moments of uranium, cobalt, and germanium, and it is mainly derived from the cobalt in UCoGe.
Figure~\ref{C-T_b} illustrates the temperature dependences of $C/T$ in several magnetic fields along the $b$ axis.
$T_{\rm SC}(H)$ or $H_{\rm c2}(T)$ are determined by the midpoint of a  $C/T$ upsurge, which are depicted as arrows in Fig.~\ref{C-T_b}.
With an increase in the magnetic field, $T_{\rm SC}(H)$ decreases; however, it appears almost unaltered between 4~T and 10~T.
It is interesting to note that in the low $T$ and high field $H$ regions, the specific heat exhibits a divergent behavior, although the  $T^{-3}$ nuclear Schottky contribution is already subtracted.
According to recent theoretical considerations\cite{machida}, rich multiple phases are suggested in the $H$-$T$ phase diagrams.
The observed upturns required to be checked in an independent experiment may be related to this prediction.

\begin{figure}
\begin{center}
\includegraphics[width=8cm]{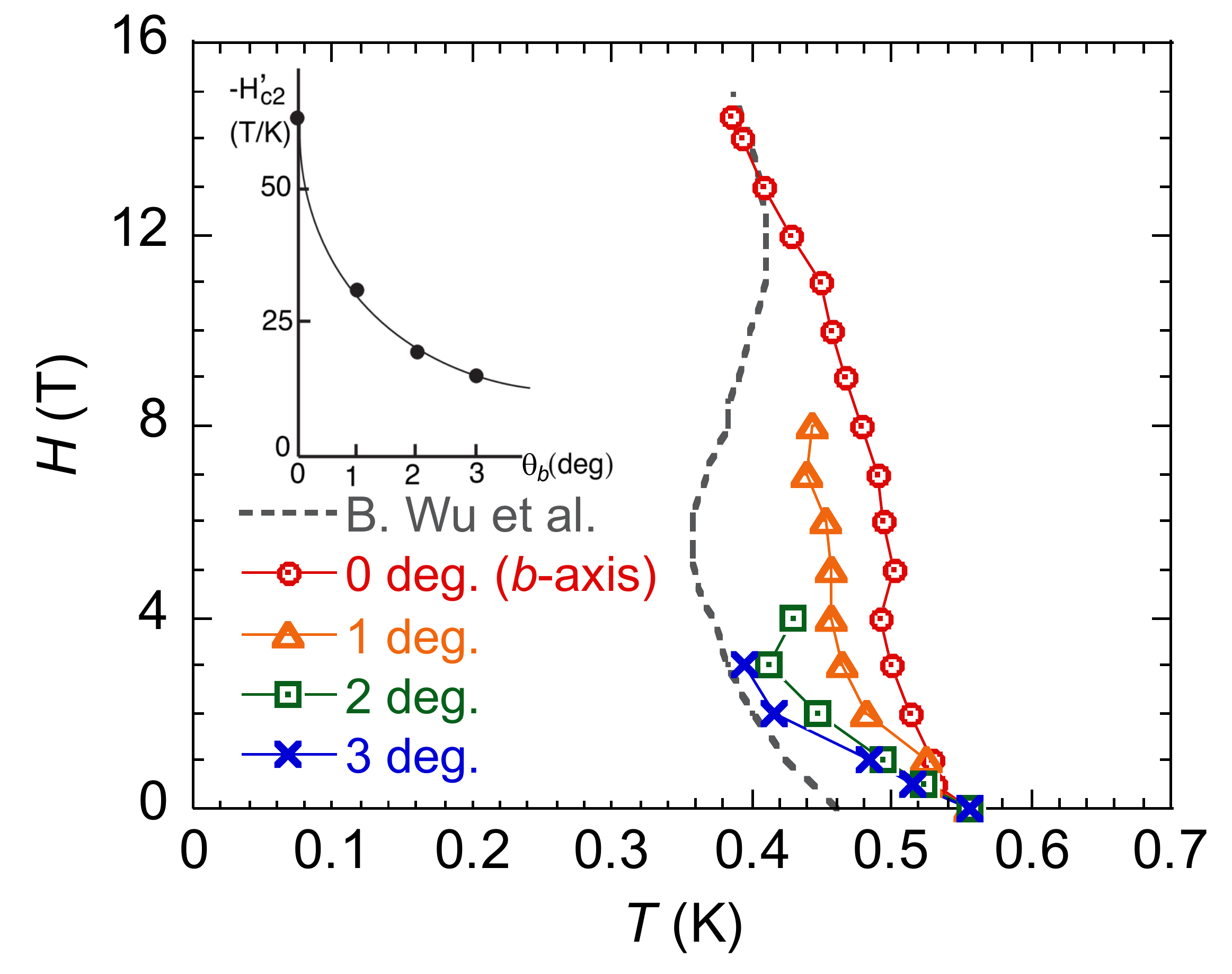}
\caption{(Color Online) $H$-$T$ phase diagram obtained at field-angles  $\theta_b = 0^{\circ}$, 1$^{\circ}$, 2$^{\circ}$, and 3$^{\circ}$ on the $bc$ plane, where $\theta_b = 0$ denotes $H \parallel b$. The inset shows the initial slopes of $H_{\rm c2}$ as a function of $\theta_b$, thereby   indicating substantial enhancements of $H_{\rm c2}$ toward the $b$ axis. }
\label{PD}
\end{center}
\end{figure}

We measured temperature dependences of $C/T$ at field-angles $\theta_b = 0^{\circ}$, 1$^{\circ}$, 2$^{\circ}$, and 3$^{\circ}$  
from the $b$ axis toward the $c$ axis, and obtained the $H$-$T$ phase diagram of UCoGe, as illustrated in Fig.~\ref{PD}. 
The phase diagram plotted together with $H_{\rm c2}$ is determined 
by the previous resistivity measurements presented in Ref.~\onlinecite{aoki2}.
The superconducting state of UCoGe  reaches above 14.5~T, along the $b$ axis ($\theta_b = 0$). 
Although an S-shaped $T_{\rm SC}(H)$  is not observed at $\theta_b = 0$ in present measurements, we can observe the  anomalous upturn behaviors of $T_{\rm SC}(H)$.
The absence of the S-shape can be attributed to the suppression of a $C/T$ jump.
In higher magnetic field regions, a peak of $C/T$ near $T_{\rm SC}(H)$ broadens, and it is difficult to precisely determine $T_{\rm SC}(H)$. 
Furthermore, the precise shape of $H_{\rm c2}(T)$ depends on experimental methods.

With an increase in $\theta_b$, the slope of $T_{\rm SC}(H)$ smoothens immediately in the lower magnetic fields below 3~T.
The inset of Fig.~\ref{PD} illustrates the initial slope changes as a function of $\theta_b$.
It can be observed that the initial slope of $H_{\rm c2}(T)$ is significantly enhanced toward the $b$ axis.
This low initial slope at $\theta_b=3^{\circ}$ smoothly approaches that of $\theta_b=90^{\circ}\parallel c$.
The abrupt increases in $H_{\rm c2}$ around the $a$ and $b$ axes that occurred just within a few degrees
are certainly anomalous and, hence, require more elucidation (see also Fig. 2 in Ref.~\onlinecite{wu}).
We will provide a number of experimental hints in the following to clarify its origin.

\begin{figure}
\begin{center}
\includegraphics[width=8cm]{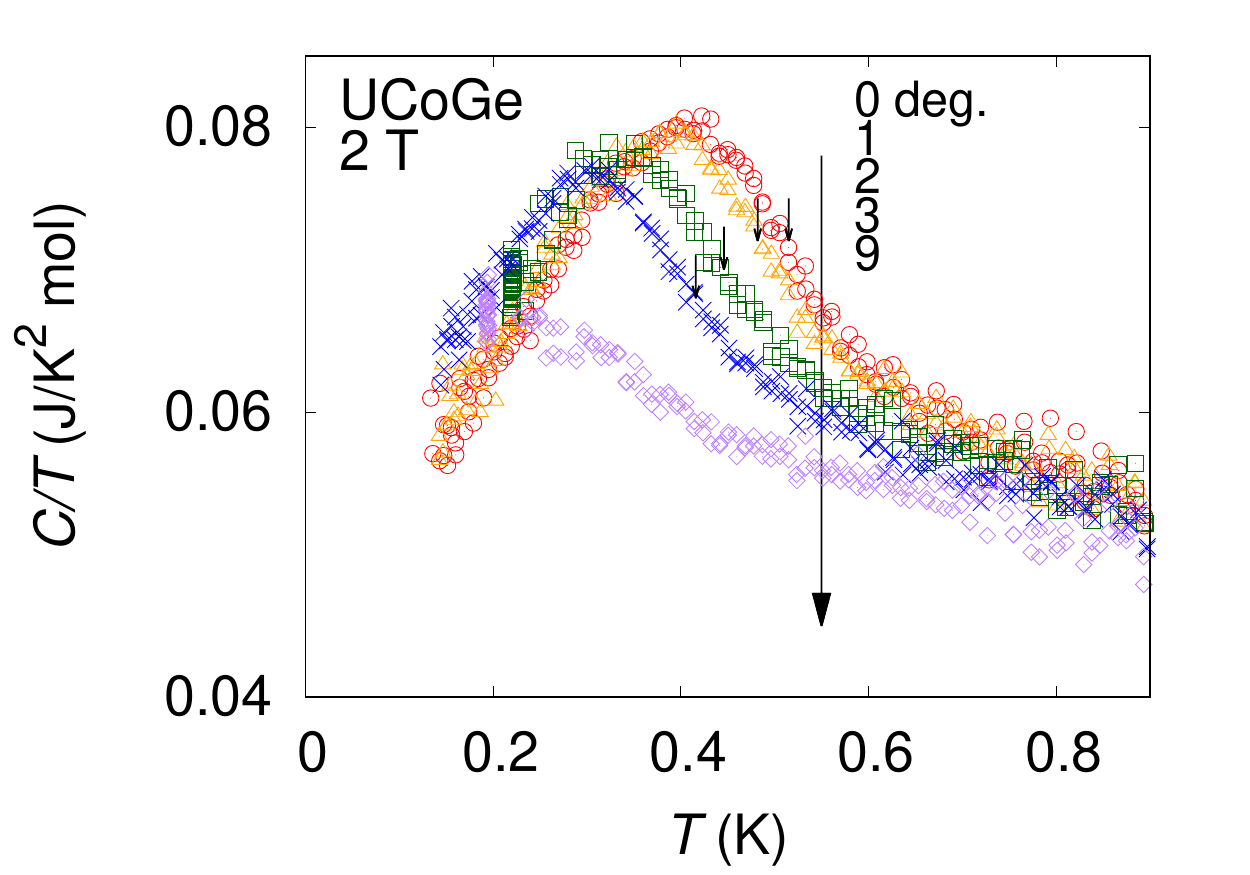}
\caption{(Color Online) Temperature dependence of $C/T$ measured at field-angles  $\theta_b = 0^{\circ}$, 1$^{\circ}$, 2$^{\circ}$, 3$^{\circ}$, and 9$^{\circ}$ on the $bc$ plane at 2~T, where $\theta_b = 0$ is $H \parallel b$.  
The arrows denote the position of  the critical temperature of the SC state $T_{\rm SC}(H)$.}
\label{angle_2T}
\end{center}
\end{figure}

Figure~\ref{angle_2T} illustrates the magnetic-field-angle $\theta_b$ variation of $C/T$ at 2~T near the $b$ axis on the $bc$ plane.
Clear specific heat jumps associated with SC transitions are observed.
$T_{\rm SC}(H)$ shifts to lower temperatures as $\theta_b$ increases, and it is smeared out at $\theta_b = 9^{\circ}$.
The arrows in Fig.~\ref{angle_2T} are denoted in Fig.~\ref{PD}.

\subsection{$\gamma(H)$ measurements near the $b$ axis}

The field-angle-dependence of $C/T$, which reflects the density of state of electrons at the Fermi level or the 
Sommerfeld coefficient $\gamma(H)$, was investigated under the $H$ applied approximately parallel to
 the $b$ axis on the $bc$ plane.

\begin{figure}
\begin{center}
\includegraphics[width=8cm]{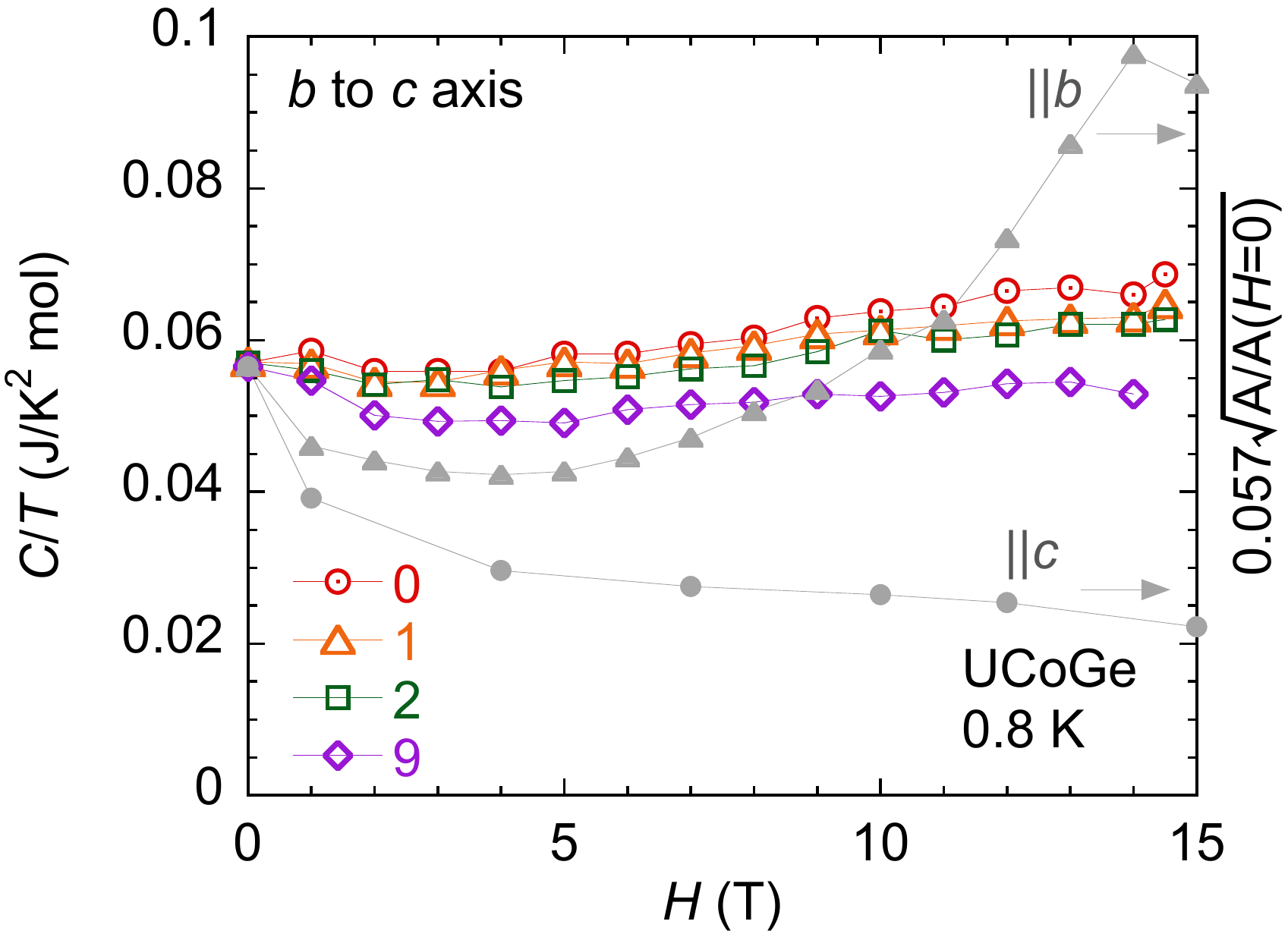}
\caption{(Color Online) Magnetic field dependence of $C/T$ measured at field-angles  
$\theta_b = 0^{\circ}$, 1$^{\circ}$, 2$^{\circ}$, and 9$^{\circ}$ on the $bc$ plane at 0.8~K  in the FM state, 
where $\theta_b = 0$ is $H \parallel b$.
The gray closed symbols are obtained from the resistivity data for$H\parallel b$ and $c$ in Ref.\onlinecite{aoki2}. To compare with $C/T$, these data are multiplied by $C/T (H = 0)$ = 0.057~J/K$^2$mol.}
\label{gamma}
\end{center}
\end{figure}

Figure~\ref{gamma} illustrates the magnetic field dependences of $C/T$ or $\gamma(H)$ measured at field-angles  
$\theta_b = 0^{\circ}$, $1^{\circ}$, $2^{\circ}$, and $9^{\circ}$ on the $bc$ plane  at 0.8~K.
At low fields, $\gamma(H^b)$ initially decreases, taking a minimum of $H^b\sim3$~T.
Upon further increase in $H^b$, $\gamma(H^b)$ starts to grow, and is greater than the $\gamma(H=0)$ value or 
$\gamma_N$ for the normal state DOS,
 exhibiting a maximum at around $H^b\sim14$~T.

In previous resistivity measurements~\cite{aoki2, aoki4}, the enhancement of $A(H^b)$ 
in higher magnetic fields along the $b$ axis was reported.
This overall field evolution of $\gamma(H^b)$ is completely consistent 
with that obtained by the resistivity measurement~\cite{aoki2,aoki4}.
The field dependence of $\gamma(H^b)$ and $\sqrt{A(H^b)}$ qualitatively coincide with each other
because of the Kadowaki-Woods relationship, where the temperature dependence of resistivity $\rho(T)$ is assumed to be
$\rho(T)=\rho_0(H^b)+A(H^b)T^2$ with $\rho_0$ being the residual resistivity.

Furthermore, the low field minimum behavior of $\gamma(H^b)$ around $H^b\sim3$T mentioned above
is consistent with the $M^b(T)$ measurement~\cite{miyakeM(T)}, and in this field region, the $M^b(T)$ curves exhibit an upward curvature,  thus implying that $\gamma(H^b)$ should decrease using the Maxwell relationship, as comprehensively explained later.
For later theoretical discussions, we point out that the enhancement of $\gamma(H^b)$  in higher fields is also observed in URhGe, and 
the enhanced field region intriguingly coincides with the reentrant $H_{\rm c2}$ region in a magnetic field along the $b$ axis~\cite{miyake,hardy,aoki4}.

When a magnetic field is tilted away from the $b$ axis toward the $c$ axis,  $\gamma(H)$ quickly decreases, 
and the enhancement of $\gamma(H)$ in higher magnetic field regions is rapidly suppressed, as illustrated in Fig.~\ref{gamma}.
This tendency continues up to the case for $H\parallel c$.
$\gamma(H^b)$ and $\gamma(H^c)$, which are inferred from $A(H^b)$ and $A(H^c)$ via resistivity measurements~\cite{aoki2}, are displayed in Fig.~\ref{gamma}.
Both $\gamma(H^b)$ and $\gamma(H^c)$ exhibits a strong suppression at low field region, and then monotonically decreases toward higher fields without  a minimum in the  $\gamma(H^c)$, or a maximum existed in  the $\gamma(H^b)$ case.
The physical reasons of these points are comprehensively discussed later from a thermodynamic perspective.

\subsection{Field angle rotating specific heat measurements}

We investigate the anisotropy of SC and FM states of UCoGe, where the the SC transition 
temperature $T_{\rm SC}(H=0) \sim 0.55$~K.
The field-angle-resolved specific heat of UCoGe was measured on the $ab$-, $ac$-, and $bc$-planes.

\begin{figure}
\begin{center}
\includegraphics[width=8cm]{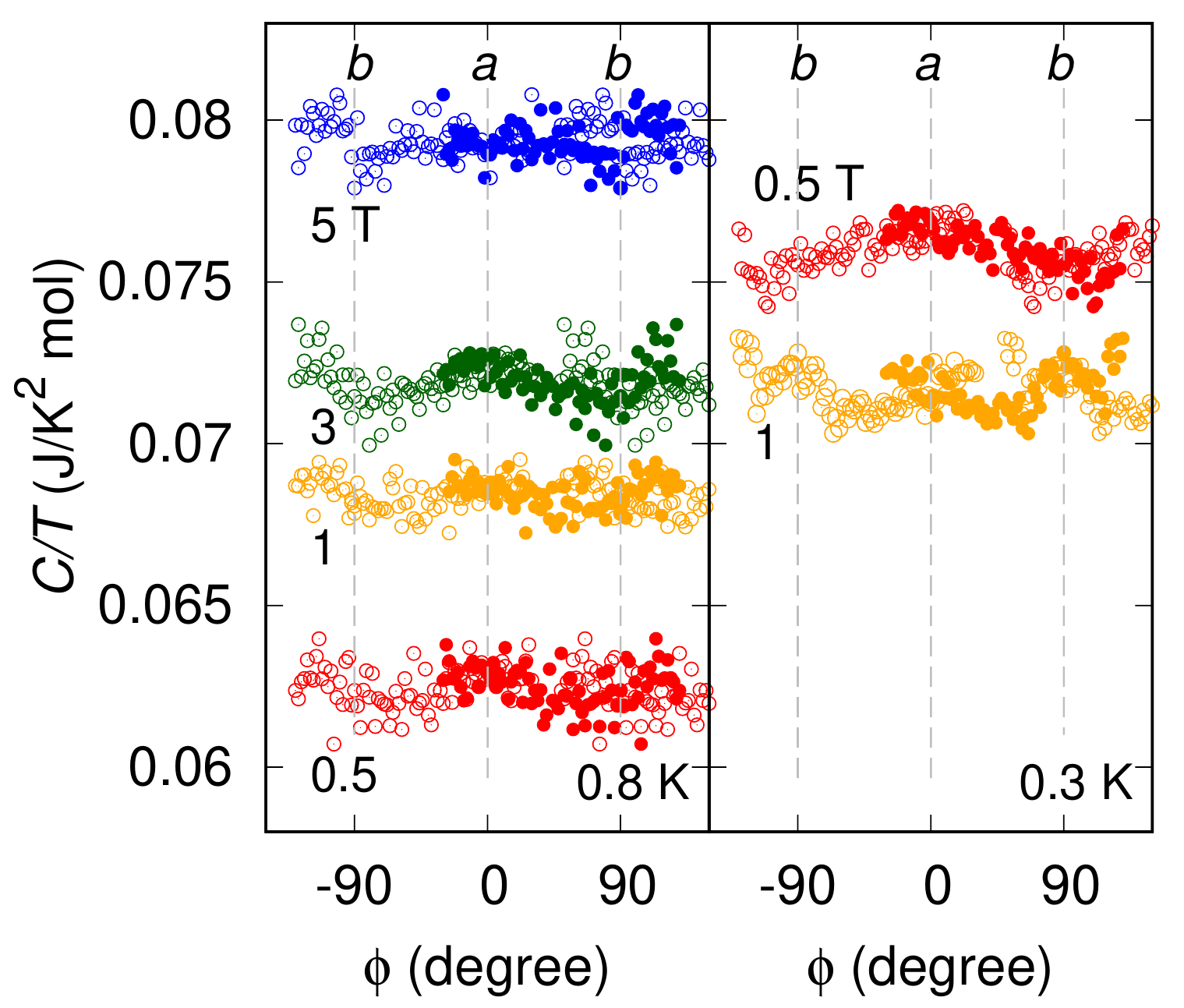}
\caption{(Color Online) Field-angle $\phi$ dependences of $C/T$ at 0.3 and 0.8 K measured in a rotating 
magnetic field within the $ab$ plane. For clarity, the data at 0.8~K are vertically shifted by 0.006 from the lowest field data of 0.5~T. Open circles represent data mirrored by the $a$ axis.}
\label{Phi2}
\end{center}
\end{figure}

Figure~\ref{Phi2} presents the magnetic-field-angle $\phi$ dependences of $C/T$ at 0.3 and 0.8 K measured in 
a rotating magnetic field within the $ab$ plane.
Small oscillations are observed both in the SC and FM states on the $ab$ plane where
$\gamma(H^a)>\gamma(H^b)$ for $H~=~0.5$~T, 1~T, and 3~T in $T~=~0.8$~K. This is consistent with the $A(H)$ data obtained via resistivity measurements~\cite{aoki2}.
At approximately $H$~=~5~T,  $\gamma(H^b)$ starts increasing, as illustrated in Fig.~\ref{gamma}.
It is evident that in the FM state, the DOS is weakly anisotropic within the $ab$ plane, thus 
implying that the DOS structure is not perfectly uniaxial, relative to the magnetic easy $c$ axis.
We also note that the oscillation patterns at $H$~=~0.5~T in the SC state at 0.3~K are remarkably similar to that
in the FM state at $H$~=~3~T and 0.8~K. The DOS structure in the SC is governed by the FM state where the
DOS in the SC is buried. 
Although in principle, the angle-resolved specific heat experiments\cite{sakaki1,sakaki2} in the SC state are beneficial in yielding 
the gap structure and the nodal position, we were unable to extract the information of the SC gap structure in this system.

\begin{figure}
\begin{center}
\includegraphics[width=8cm]{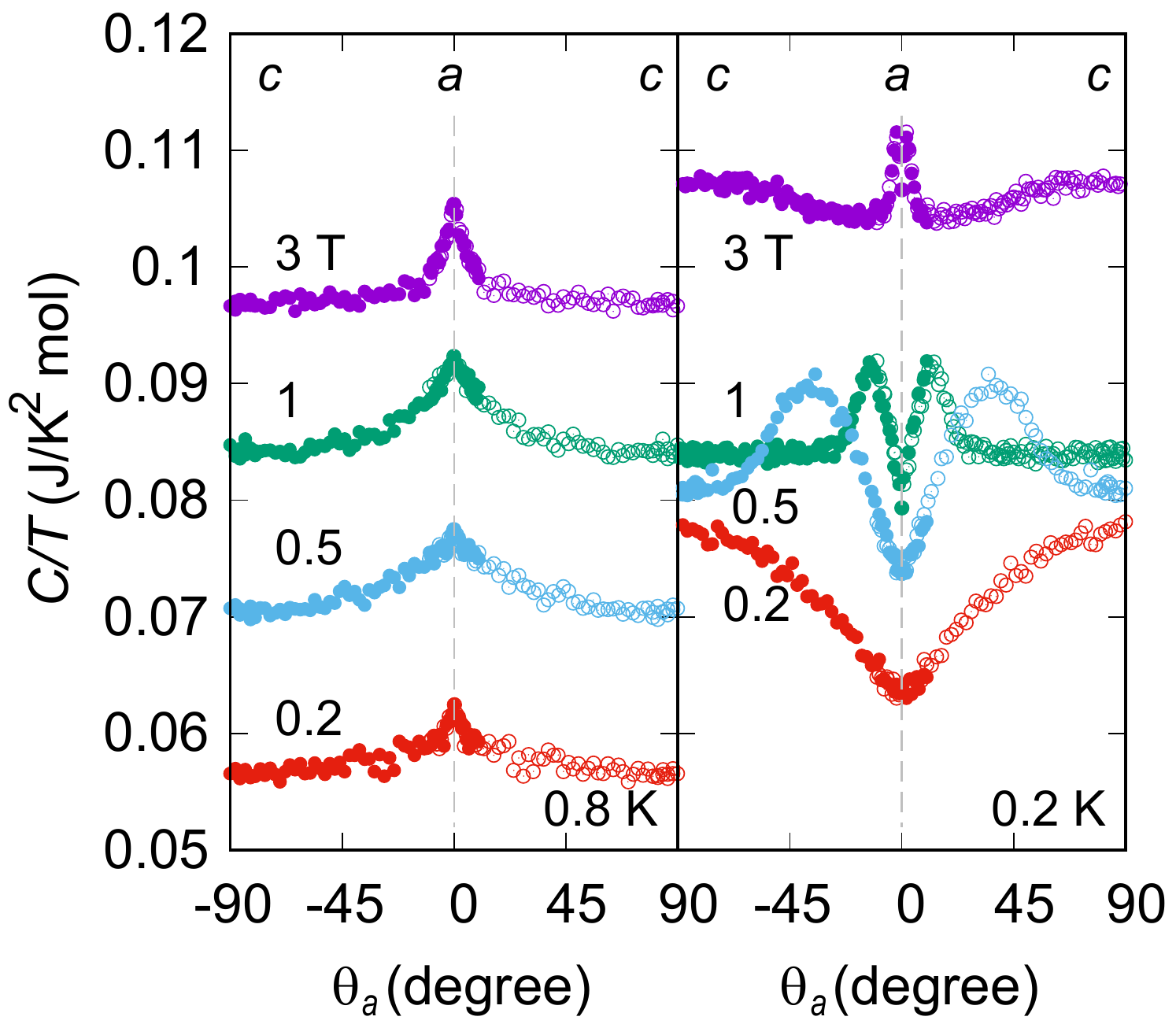}
\caption{(Color Online) Field-angle $\theta_a$ dependences of $C/T$ at 0.2~K and 0.8 K measured in a rotating magnetic field within the $ac$ plane. For clarity, the data are vertically shifted by 0.01 and 0.015 at 0.2 and 0.8 K from the lowest field data of 0.2~T, respectively.  Open circles represent the data mirrored by the $a$ axis.}
\label{THsweep_a}
\end{center}
\end{figure}

\begin{figure}
\begin{center}
\includegraphics[width=8cm]{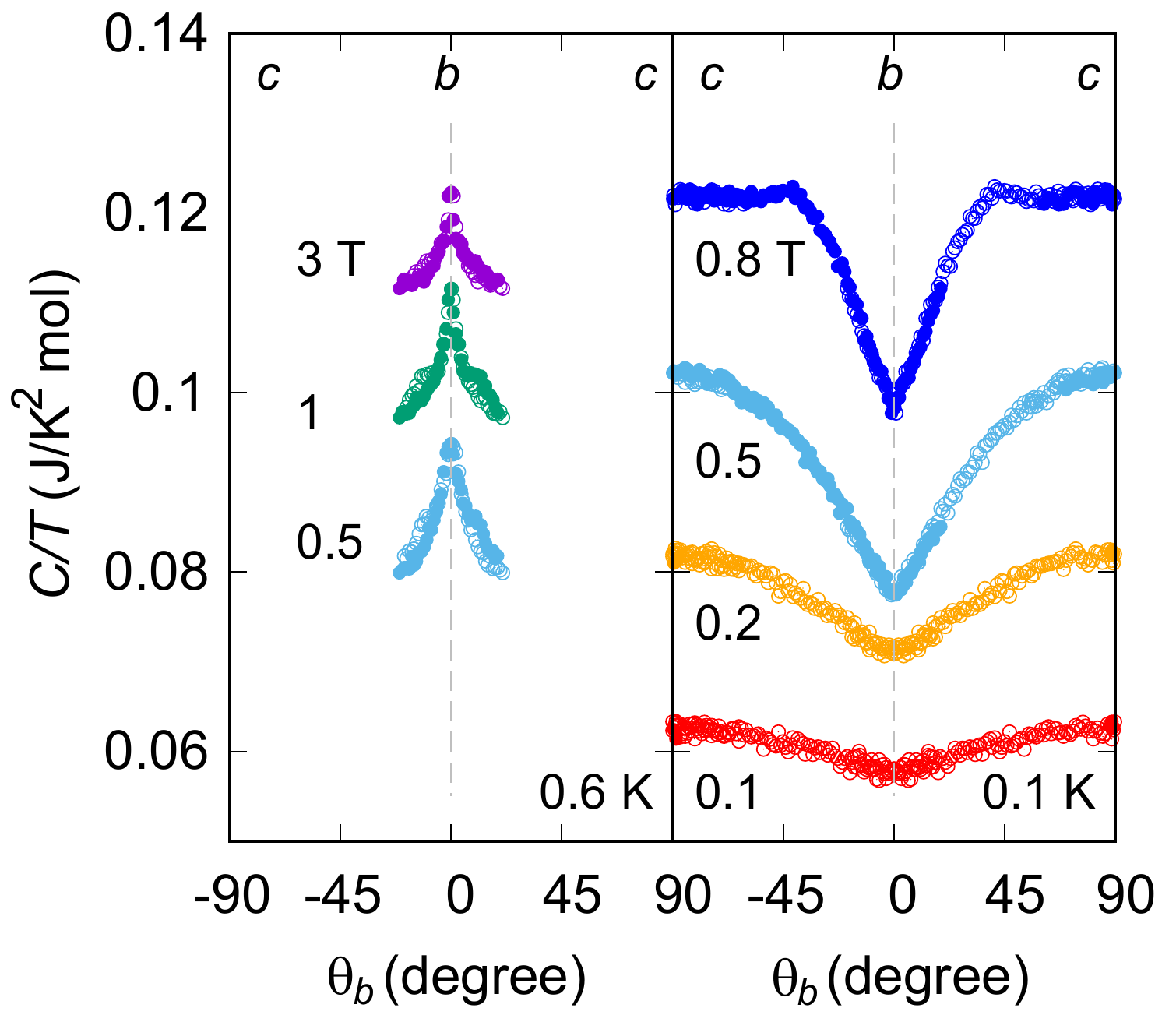}
\caption{(Color Online) Field-angle $\theta_b$ dependences of $C/T$ at 0.1~K and 0.6 K measured in a rotating magnetic field within the $bc$ plane. For clarity, the data are vertically shifted by 0.015 and 0.02 at 0.1 and 0.6 K from the lowest field data, respectively.  Open circles denote the data mirrored by the $b$ axis.}
\label{THsweep_b}
\end{center}
\end{figure}

Figures~\ref{THsweep_a} and \ref{THsweep_b}  present the field-angles $\theta_a$ and $\theta_b$ dependences of $C/T$ measured in a rotating magnetic field within the $ac$ and $bc$ planes, respectively. 
In the FM state, $C/T$ is significantly suppressed when the magnetic field is tilted away from the $ab$ plane toward the $c$ axis.
It takes a maximum value at the $a$ or $b$ axis, which is approximately 10$\%$ larger than that of the $c$ axis.

The anisotropic behaviors inherent in the SC state shown in Figs.~\ref{THsweep_a} and \ref{THsweep_b}  are qualitatively the same as that in the FM state, thus indicating that the Ising anisotropy of the FM state is dominant, even in the SC state.
The Ising-like magnetic  anisotropy masks the gap structure of UCoGe.
Here, the double peaks observed in a magnetic field of 0.5 and 1 T at 0.2~K on an $ac$ plane reflect the phase transition from the FM state to the SC state.
The peaks shift to a smaller $\theta_a$ and  $\theta_b$ range in higher magnetic fields, because $H^c$ rapidly becomes larger as $\theta_a$ and  $\theta_b$ increase.
In lower magnetic fields, the peaks do not appear because the SC state is not broken.
As observed from the data in the SC at 0.1~K in Fig.\ref{THsweep_b},  $\gamma(H^c)>\gamma(H^b)$ solely because 
$H_{\rm c2}^b>H_{\rm c2}^c$, where the possible nodal structure is hidden.

\begin{figure*}
\begin{center}
\includegraphics[width=16cm]{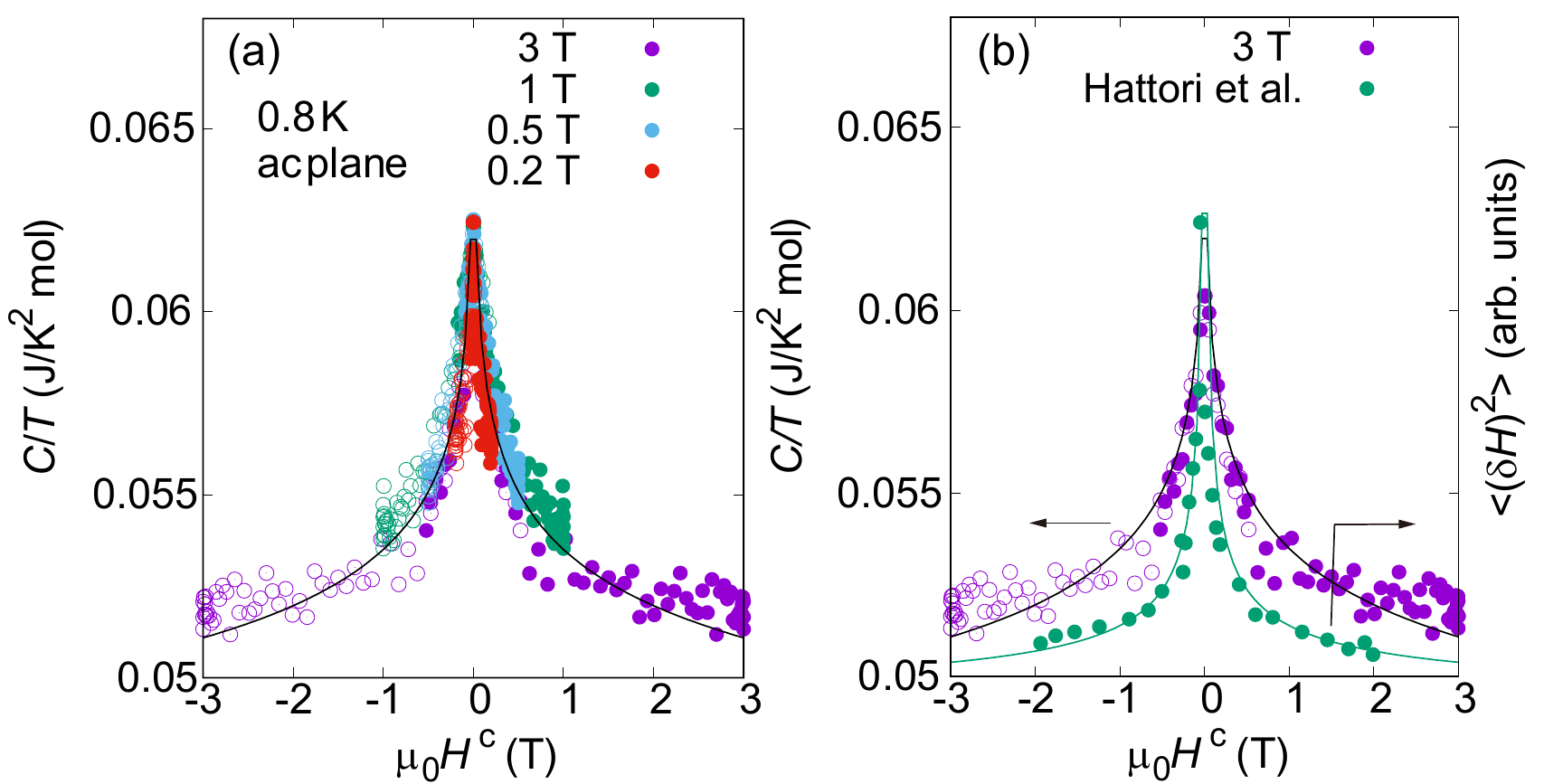}
\caption{(Color Online) (a) $C/T$ vs $H^c$ plot of the data shown in Fig.~\ref{THsweep_a}, 
 which were measured at 0.8~K on the $ac$ plane. A solid line denotes a function that is proportional to $(H^c)^{-0.04}$. 
Open circles represent the data mirrored by the $a$ axis. (b) Comparison between $C/T$ vs $H^c$ plot at 3~T in figure~8(a) and longitudinal spin fluctuation $<(\delta H)^2>$ calculated from 1/T$_1$ at 3.5~T and 1.7~K in previous NMR measurements~\cite{hattori}.}
\label{Hc}
\end{center}
\end{figure*}

Figure~\ref{Hc}(a) illustrates the $C/T$ vs $H^c$ plot of the data presented in Fig.~\ref{THsweep_a}, which were measured at 0.8~K on the $ac$ plane. 
Here, the data on the $bc$ plane is omitted because almost the same result was obtained.
$C/T$ is logarithmically suppressed  ($\propto (H^c)^{-0.04}$) as $H^c$ increases, thus reflecting the Ising-like magnetic anisotropy.
Similar logarithmic suppressions of $1/T_1(H^c)\propto {1/\sqrt {H^c}}$ and $H_{\rm c2}$ are also reported in a previous study~\cite{hattori}.
As shown in figure~\ref{Hc}(b), Hattori et al. attributed the strong $1/T_1(H^c)$ change to the suppression of longitudinal spin fluctuations.
We will comprehensively discuss these points later.

\subsection{Magnetization measurements}

Up to this point, we have concentrated on the specific heat experiments in various aspects to extract the 
DOS structures in the FM and SC states. In particular, we deduced the strong suppression of the DOS
toward the $c$ axis, as shown in Fig.~\ref{Hc}.
To investigate its origin, we are going to proceed with the $M^c(T)$ in this subsection.

\begin{figure}[t]
\includegraphics[width=8cm]{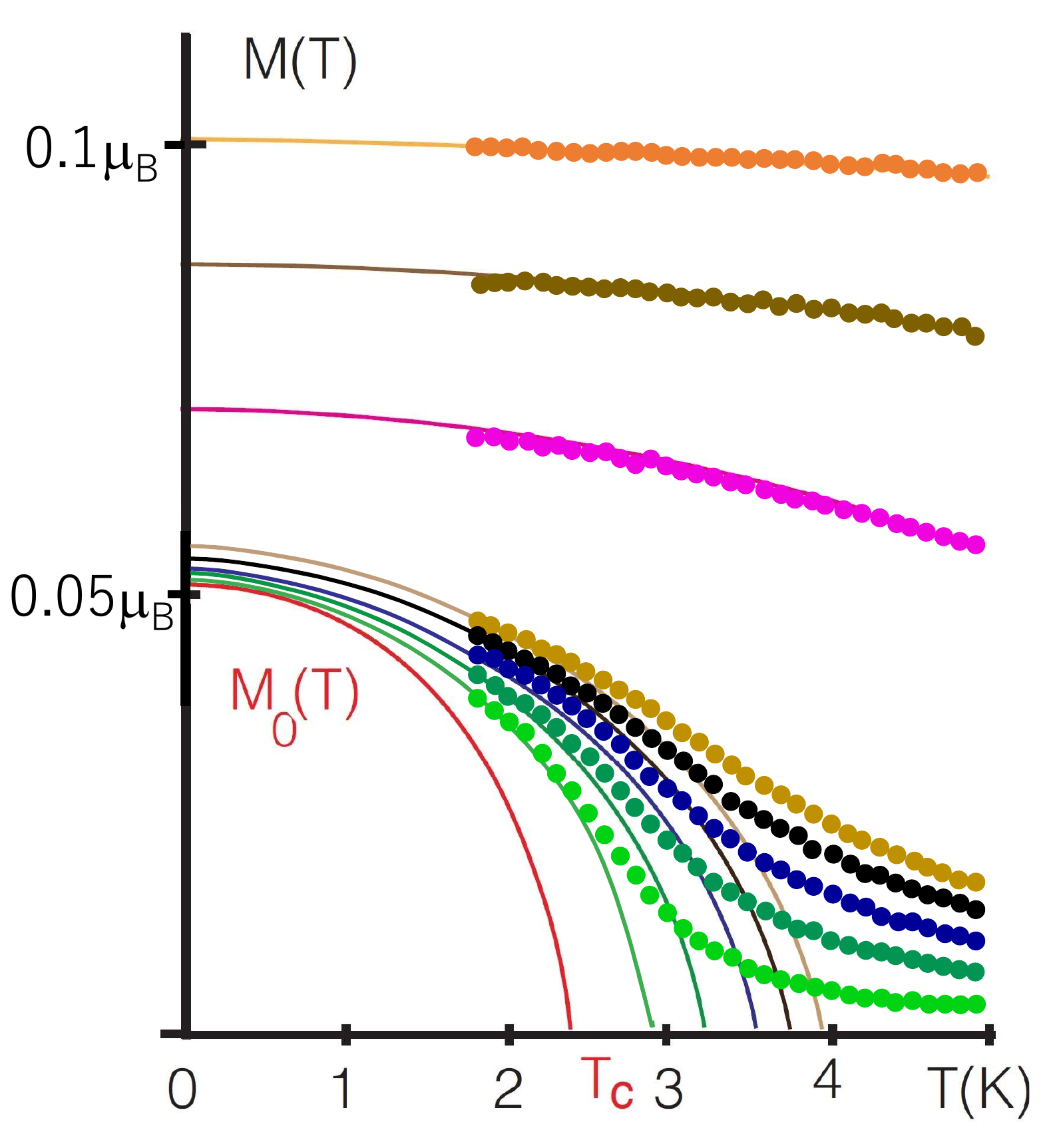}
\caption{(Color Online) Temperature dependences of the magnetization $M(T)$ for $H\parallel c$  in
various field strengths. The dots denote the experimental data where
$H^c$~=~0.02, 0.04, 0.06, 0.08, 0.1, 0.5, 1.0, and 1.5~T from the bottom.
Here, we also plot the approximate spontaneous moment curve $M_0(T)$ with $T_C=2.5$~K
and $M_0(T=0)=0.05\mu_B$/U with red color. The continuous lines denote $M(T)$ parabola curves  obtained from Eq. (3).}
\label{M(T)}
\end{figure}

Figure~\ref{M(T)} presents the data of $M^c(T)$ for $H\parallel c$ measured by using MPMS at low temperatures, together with the spontaneous moment
curve $M_0(T)$ without a field.  $M_0(T)$ is obtained by smoothly connecting $M_0(T=0)=0.05\mu_B/$U and the Curie temperature
$T_C=2.5$~K with a parabola. This yields $M_0(T)=M_0+\beta_0T^2$ with $\beta_0=-M_0/T_C^2$.
It is deduced from Fig.~\ref{M(T)} that the $M(T)$ curves smoothly and continuously evolve from the spontaneous moment
curve $M_0 (T)$ to
$M(T)$ for finite fields. Accordingly, we infer that $M_0(T)$ is a basis for elucidating the $M(T)$ curves at low fields.
In other words, once the spontaneous moment and $T_C$ are known, $\beta(H)$ can be deduced at low fields.

In general, according to the Maxwell relation, $\partial M/\partial T = \partial S/\partial H$ then,

\begin{eqnarray}
{d\over dH}{C\over T}={\partial^2\over\partial T^2}M(T,H),
\label{maxwell}
\end{eqnarray} 

\noindent
in the low $T$ limit,

\begin{eqnarray}
{d\gamma(H)\over dH}=2\beta(H),
\label{maxwell1}
\end{eqnarray}
 
 \noindent
where we have introduced $\beta(H)$ by phenomenon theory

\begin{eqnarray}
M(T)=M_0+\beta(H)T^2.
\label{beta(H)}
\end{eqnarray}

\noindent 
This relationship is well satisfied by the data shown in Fig.~\ref{M(T)}, which implies that we can extract $\beta(H^c)$ reliably.

In this low field region $\beta(H^c)<0$, because the parabolas point upward and flatten as $H$ increases, as illustrated in Fig.~\ref{M(T)}. This implies $d\gamma(H^c)/dH^c~<~0$, i.e. $\gamma(H^c)$ decreases as a function of $H^c$.
Eventually, as $H$ increases, $\beta(H^c)$ tends toward zero from below.
As comprehensively discussed below, this is quite a general property that under the applied field parallel to the magnetic easy axis, the 
DOS first decreases $H^c$ linearly, and then becomes flat. This is true for URhGe~\cite{hardy}.

\begin{figure}
\includegraphics[width=8cm]{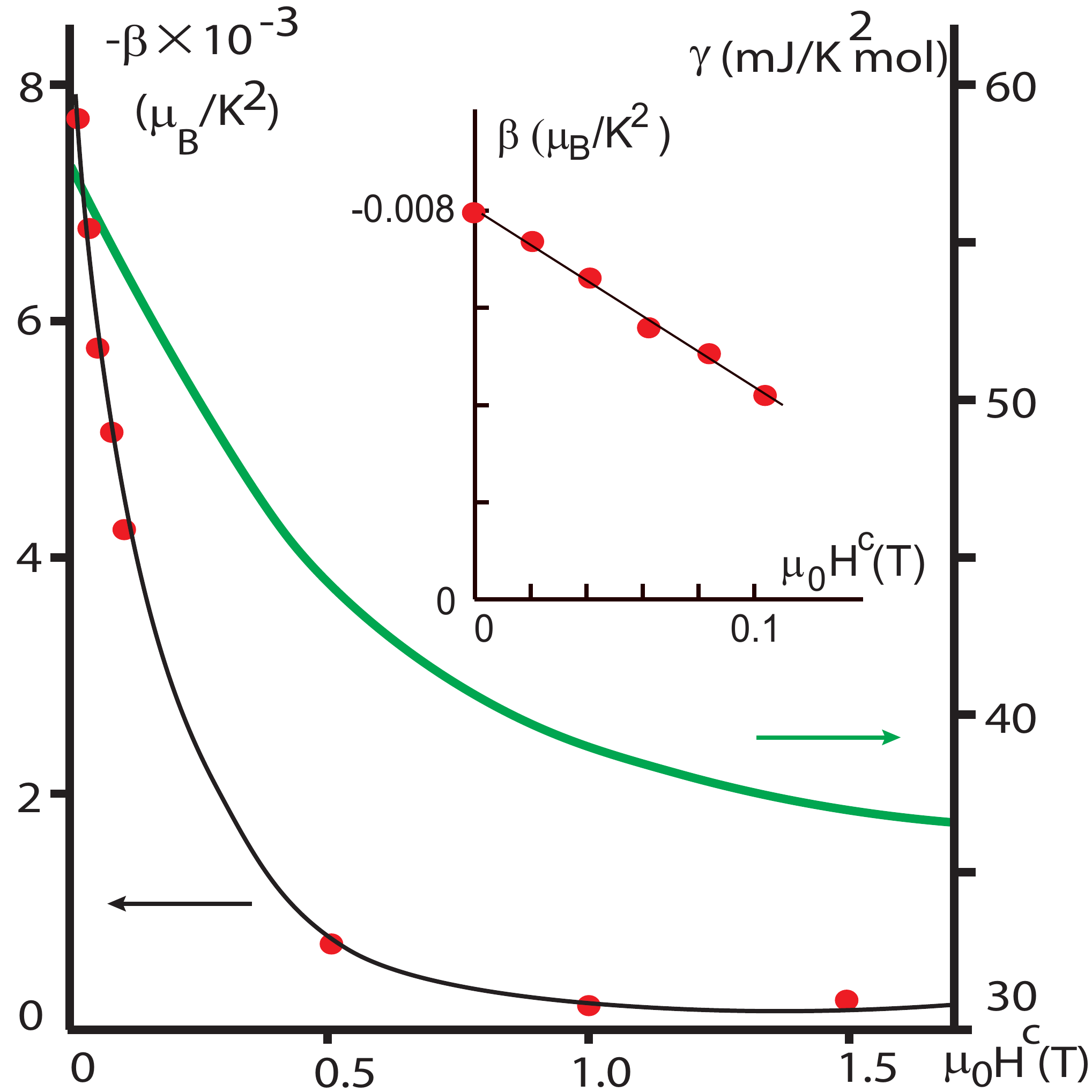}
\caption{(Color Online) By using the data in Fig.~\ref{M(T)}, $\beta(H^c)$  values are evaluated.
We assume that the $M(T)$ curves centered at $T=0$ are approximated by parabolas.
The value at $\beta(H=0)$ is evaluated by $\beta_0=-M_0/T_C^2$.
The smooth curve expresses the fitting with $\beta(H^c)\propto-(H^c)^{-3/2}$
for 0.3~T~$<~H^c~<~$1.5~T. 
From $\beta(H^c)$, $\gamma(H^c)$ (green curve) is evaluated.
The inset shows the linear behaviors  of $\beta(H^c)$ in $H^c$ up to $H^c~=~$0.1~T.
}
\label{beta}
\end{figure}

Using the data presented in Fig.~\ref{M(T)}, we can estimate $\beta(H^c)$ and  $\gamma(H^c)$.
The obtained results are presented in Fig.~\ref{beta}. 
$|\beta(H^c)|$ abruptly decreases and approaches zero from $H^c=0$ to $H^c~=$~1.5~T, as demonstrated in the figure.
The linearity of $\beta(H^c)$ in $H$ at low fields is observed from the inset of Fig.~\ref{beta}.
The finite field estimates of $\beta(H^c)$ are determined to smoothly tend to the estimate by $\beta_0=-M_0/T_C^2$,
thus implying that the $\beta_0$ value is reliable, and the resulting initial slope of$\gamma(H^c)$ near $H^c=0$ must be
compatible with the experimental data. 

In fact, according to the Maxwell relation expressed in Eq.~(\ref{maxwell1})
with $\beta(H)=\beta_0+\beta_1H$ for $H^c<0.1$~T with $\beta_0=-0.008\mu_B$/K$^2$,
we obtain

\begin{eqnarray}
\gamma(H)&=&\gamma_N+2\int_0^H\beta(H)dH \nonumber\\
&=&\gamma(0)+2\beta_0H+\beta_1O(H^2)
\label{gamma0.1T}
\end{eqnarray}

\noindent
$\gamma(H^c)$ drops to 57~mJ/${\rm K}^2$mol from $\gamma(0)$ by the amount of 7~mJ/${\rm K}^2$mol at $H^c~=~0.1$~T.
As observed from Fig.~\ref{Hc}, this is precisely matched with the peak height at that point.

For 0.3~T$~<~$$H^c$$~<~$1.5~T it is deduced that $\beta(H^c)\propto -(H^c)^{-3/2}$, thus upon integration in this field region,

\begin{eqnarray}
\gamma(H^c)\propto (H^c)^{-1/2}
\label{gamma0..3T}
\end{eqnarray}

\noindent
This singular behavior is observed in Fig.~\ref{Hc}.
It is noted that the precise power index depends relatively on the fitting field regions. 
In fact, according to the NMR experiment~\cite{hattori},  $1/T_1\propto (H^c)^{-1/2}$, which slightly differs from the expected exponent -1; however, both quantities $\gamma(H^c)$ and $1/T_1$ prove the same physical phenomenon toward the $c$ axis.

In Fig. \ref{beta}, we display $\gamma(H^c)$ thus obtained as the green curve, showing that $\gamma(H^c)$ decreases sharply with $H^c$.
This behavior is consistent with $\sqrt{A(H^c)}$ by the resistivity measurement \cite{aoki2}.

\section{Theoretical considerations}

\subsection{$\gamma(H)$ along the magnetic easy axis}
It is important to understand the common experimental fact under the applied field along the magnetic easy axis: three compounds, UCoGe~\cite{aoki2}, URhGe~\cite{hardy}, and UTe$_2$~\cite{aokiUTe2,miyakeUTe2}, exhibit a decrease in $\gamma(H^c)$.
This is understood for the FM ordered state in terms of the $M(T)$ behaviors.
Because the spontaneous FM moment $M_0(T)$ as an order parameter of itinerant ferromagnets is altered like a BCS form, i.e., near $T~=~0$,

\begin{eqnarray}
M_0(T)=M_0+\beta_0 T^2+O(T^4)
\label{M_0(T)}
\end{eqnarray}

\noindent
with $\beta_0<0$, quite generically.
Hence, in lower field regions, $\gamma(H)$ decreases via the Maxwell relation expressed in Eq.~(\ref{maxwell1}), as described by Eq.~(\ref{gamma0.1T}): $\gamma(H^c)=\gamma(0)+2\beta_0H^c$.
This implies that 

\noindent
(1) $\gamma(H^c)$ for the field applied magnetic easy axis should always decrease linearly in $H$. 

\noindent
(2) The linear slope is determined by the value of $\beta_0$ without a field.

\noindent
(3) A beneficial estimate of  $\beta_0$ is given by $\beta_0=-M_0/T_C^2$, which is obtained by
regarding $M_0(T)$ as a parabola.

These considerations provide a general formula for $\gamma(H^c)$, which is given by

\begin{eqnarray}
 \gamma(H^c)=\gamma(0)-2{M_0\over T_C^2}H^c.
\label{gamma0}
\end{eqnarray}

\noindent
Remarkably, the initial change in  $\gamma(H^c)$ for the magnetic easy axis is 
$H$-linear and determined by the zero-field information $\beta_0$.
This fact is universal for itinerant ferromagnets, independent of the underlying material parameters.
Although we do not know the microscopic origins of this decrease, it is physically plausible that because the applied field along the easy axis directly destabilizes the optimized FM state at the zero field
by forcing the spontaneous moment $M_0$ to increase, the DOS may decrease.
For UCoGe, $M_0=0.05\mu_B/$U and $T_C=2.5$~K, $\beta_0({\rm UCoGe})=-0.008\mu_B/{\rm K}^2$ whereas
for URhGe, $M_0=0.4\mu_B/$U and $T_C=9.5$~K, $\beta_0({\rm URhGe})=-0.004\mu_B/{\rm K}^2$.
This means that the initial decrease in  $\gamma(H^c)$ in UCoGe is as steep as or steeper than that in URhGe.
This approximately coincides with an already established experimental fact~\cite{hardy,aokireview}.
UTe$_2$ is intriguing because, although there is no static long-ranged FM found in this compound~\cite{aokiUTe2,miyakeUTe2,kittaka}.


\subsection{$\gamma(H^b)$ along the $b$ axis}

Because the applied field along the hard axis $b$ is interesting, as it reinforces upper critical field of SC $H^b_{\rm c2}$ along the $b$ axis shown in Fig.~\ref{PD}, it is worth examining $\gamma(H^b)$ comprehensively.
Figure~\ref{gamma} shows a typical pattern for the common field evolution of $\gamma(H^b)$.
At low fields, $\gamma(H^b)$ is slightly decreases, gradually increases toward higher fields, and then exhibiting a maxmum at characteristic peak fields $H^b_{peak}$.
Then after exhibiting a maximum at certain characteristic field, $\gamma(H^b)$ decreases further in higher fields.
$H^b_{peak}$ indicates the quantum critical point with $T_C(H^b)\rightarrow0$ for UCoGe.
The quantum criticality is responsible for $\gamma(H^b)$ to exhibit a peak.

Although no theory succeeds in explaining it microscopically, 
we can understand it  phenomenologically in terms of the $M^b(T)$ curves via the Maxwell relation expressed in Eq.~(\ref{maxwell1}):
At lower fields, $M^b(T)$ exhibits a global maximum at  $T_C(H^b)$ by definition, thus 
$\beta(H^b)$ is positive, or $\gamma(H^b)$ increases. On increasing $H^b$,
 $T_C(H^b)$ decreases and simultaneously  $\beta(H^b)$ becomes larger.
This means that $\gamma(H^b)$ is further enhanced  toward $H^b\rightarrow H^b_{peak}$ from below.
Above  $H^b>H^b_{peak}$, because the downward parabola of $M(T)$ changes to the upward direction at $H^b=H^b_{peak}$.
At this point, the $M^b(T)$ curves become a monotonous decreasing function of $T$
with its maximum at $T=0$. $\beta$ alters its sign from positive to negative, and $\gamma(H^b)$ starts to decrease.

The $b$ axis is also the magnetic hard axis for URhGe, and UTe$_2$.
The applied field along the hard axis $b$ induces the reentrant SC in URhGe, it is worth examining $\gamma(H^b)$ comprehensively.
Field evolution of $\gamma(H^b)$ of these two compounds are similar to that of UCoGe.
At low fields, $\gamma(H^b)$ is almost constant  (URhGe~\cite{hardy}, UTe$_2$~\cite{kittaka,miyakeUTe2}), gradually increases toward higher fields.
Then after exhibiting a maximum at certain characteristic  peak field. 
The characteristic peak fields are different from UCoGe in nature.
In URhGe, $H^b_{peak}\sim$14~T precisely corresponds to the field at which the FM moment rotates from the
$c$ axis to the $b$ axis. 
This rotation is characterized by the weak first order transition~\cite{levy,nakamuraURhGe}.
In UTe$_2$, $H^b_{peak}=$35~T corresponds to the strong first order transition, where
the induced moment $M\parallel b$ exhibits a jump.

 \subsection{Possible second transition at low temperatures}
 
 As illustrated in Fig.~\ref{C-T_b}, in the data at $H^b$~=~10~T and 14~T, $C/T$ exhibits anomalous
 upturn at the lowest temperatures. Because the nuclear contribution is already discarded from the
 raw data, they are expected to be the intrinsic electronic specific heats.
 If true, the anomalous increases in $C/T$ may indicate a second or first order phase transition
 at further lower $T$ values, that are yet to be discovered, although it is possible that these increases are derived from impurities or background subtraction. In fact, when the pairing symmetry is a spin triplet 
 non-unitary $A_1$ phase realized in this system, the second phase transition from $A_1$ to $A_2$ phases in SC state under a magnetic field is predicted~\cite{machida}.
 This prediction is supported by the thermal conductivity measurement (see Fig.5 in Ref.~\onlinecite{wu}), which demonstrates
 a thermal-conductivity jump at $H^b\sim15$T at low $T$ values. Hence, this point deserves further investigation.
 
 \subsection{Sharp peak structures of DOS}
 
As shown in Fig.~\ref{Hc}, the observation of the sharp peak structures of the DOS about the principal $a$ and $b$ axes 
is one of the highlights of this study. The DOS or effective mass abruptly increases toward the  $a$ and $b$ axes
from the $c$ axis. This experimental fact is fully and quantitatively supported by magnetization measurements shown in 
Fig.~\ref{M(T)}.
This sharp peak structure is reminiscent of the $H_{\rm c2}$ curve displayed in the inset of Fig.~\ref{PD},
where  $H_{\rm c2}$ is significantly and abruptly enhanced toward the $b$ axis from the $c$ axis.
Therefore, it is intuitive to consider that the $H_{\rm c2}$ enhancement partly originates from the DOS increase.
 In fact,  if a BCS-type $T_{\rm SC}$ expression is assumed, $T_{\rm SC}\propto \exp^{-1/gN(0)}$ with $g$ is
 the attractive interaction. A mere increase in DOS from 0.055 to 0.065 (J/K$^2$mol) or 10$\%$ increase of DOS
 may significantly contribute to increasing $H_{\rm c2}$. Conversely, the boosted $H^b_{\rm c2}$ substantially exceeded 16~T,
 whose origin was described in terms of the $A_1$-like phase~\cite{machida},  rapidly decreases owing to the DOS changes.
 We point out that a similar sharp peak structure of $H^b_{\rm c2}$, when  positioned away from the $b$ axis, is observed in
 UTe$_2$ (refer to inset of Fig. 22 in Ref.~\onlinecite{machida}). The physical origin of these intriguing phenomena
 may be same.
 
\subsection{Pairing symmetry}

The attempt to identify the gap or possible nodal structure in UCoGe via the angle-resolved 
specific heat experiment is inhibited by the presence of the FM order above $T_{\rm SC}$,
which hides the detection of possible nodes, as illustrated in Figs.~\ref{Phi2}, ~\ref{THsweep_a}, and ~\ref{THsweep_b}.
To facilitate future experiments to detect the nodal structure in UCoGe,
we theoretically summarize the possible pairing states using classified-group theory~\cite{ohmi,annett,machida} and propose experiments.
Under the orthorhombic crystals, the allowed pairing functions belong to one-dimensional representations, both
for singlet and triplet cases. Focusing on the triplet pairing, the nodal structure is represented by line  and point nodes in weak and strong 
spin-orbit coupling cases, respectively. Because the line node structure is suggested~\cite{aokireview}, 
the former is the case if the line nodes are confirmed in future.
Then, under the FM molecular field, the Cooper pairs are polarized along the FM moment, and
the resulting pair symmetry must be non-unitary triplet symmetry~\cite{nonunitary} with line nodes.
To precisely identify the pairing symmetry, we need to determine the orientation of
the line nodes in the reciprocal space. One of the best method is the angle-resolved thermal 
conductivity measurement, which may circumvent the DOS anisotropy due to the FM state.

\subsection{Possible pairing mechanism}

Because the mass enhancement around $H^b=H^b_{peak}$ coincides with the $H^b_{\rm c2}$ enforcements
in three compounds in common, the S-shaped $H^b_{\rm c2}$ in UCoGe presented in Fig.~\ref{PD}, 
reentrant $H^b_{\rm c2}$ in URhGe, and $T_{\rm SC}(H^b)$ maximum in UTe$_2$,
it is tempting to consider that the pairing mechanism is related to the origin of this mass enhancement.
Hattori et al.~\cite{hattori} claim that because 1/$T_1$ exhibits a sharp peak structure around the $b$ axis in UCoGe,
and its shape is similar to the angle dependence of $H_{\rm c2}$ (refer to Fig.4 in Ref.~\onlinecite{hattori}),
the longitudinal spin fluctuations due to the Ising nature of the FM moment system is the origin
of the pairing mechanism because longitudinal fluctuations should be suppressed by $H^c$. 
We point out here that the sharp 1/$T_1$ peak structure itself originates from
the mass suppression, as we already demonstrated in this study. A question is how it comes. Therefore, at this moment
there is no direct information or support to conclude the pairing mechanism.
We note that according to Tokunaga et al~\cite{tokunaga}, the 1/$T_2$ is enhanced
precisely at the reentrant $H^b_{\rm c2}$ region in URhGe, which implies that the transverse spin
fluctuations relative to the easy $c$ axis are enhanced there. This is a counter example for the above scenario
under the plausible assumption that UCoGe, URhGe, and possibly UTe$_2$ are governed by the same mechanism.

\section{Conclusion and summary}

The specific heat of the uranium FM superconductor UCoGe was measured at low temperatures under high-precision angle-resolved magnetic fields along the $ab$-, $ac$-, and $bc$-planes.
The field-angle-dependence of the $H$-$T$ phase diagram was obtained in the magnetic field near the $b$ axis on the $bc$ plane.
A significant and sharp enhancement of $C/T$ was observed in the high magnetic field along the hard $b$ axis 
where the SC state reaches a high magnetic field region.
When the magnetic field was slightly tilted away from the $b$ axis toward
 the easy $c$ axis, the slope of $T_{\rm SC}(H)$ was smoothened, and $C/T$ was strongly suppressed owing to the Ising anisotropy.
The Ising anisotropy of $C/T$ in the FM state was dominant even in the SC state.
Magnetization measurements along the $c$ axis were also conducted to check the thermodynamic consistency of
the density of state enhancement.

We theoretically discussed and argued the origins of the anomalous anisotropic specific heat results along with magnetization results
and provided a perspective on the common and different features of UCoGe, URhGe, and UTe$_2$ to
further facilitate deeper understandings of our findings, which might lead to elucidating the pairing symmetry and pairing mechanism.

In summary,

\noindent
(1) The DOS suppression toward the principal $c$ axis from the $a$ and $b$ axes is abnormally sharp.

\noindent
(2) We established a generic formula of $\gamma(H)$ for $H$ parallel to the magnetic easy axis, which always decreases at low fields.


\section*{Acknowledgments}  
We greatly thank Y. Homma for his technical support on the used uranium samples.
One of the authors (KM) thanks K. Ishida and A. Miyake for many constructive discussions.
The present work was supported in part by a Grant-in-Aid for Scientific Research on Innovative Areas ``J-Physics'' (15H05883, 18H04306)  and KAKENHI (15H03682, 17K05553, 18H01161, 21K03455) from MEXT.

\end{document}